# Microemulsification: An Approach for Analytical Determinations

Renato S. Lima,[†,‡,∥] Leandro Y. Shiroma,[†,‡] Alvaro V. N. C. Teixeira,[§] José R. de Toledo,[§] Bruno C. do Couto,[⊥] Rogério M. de Carvalho,[⊥] Emanuel Carrilho,[∥,¶] Lauro T. Kubota,[*,‡,∥] and Angelo L. Gobbi[†].

[†]Laboratório Nacional de Nanotecnologia, Centro Nacional de Pesquisa em Energia e Materiais, Campinas, São Paulo 13083-970, Brasil.
[‡]Instituto de Química, Universidade Estadual de Campinas, Campinas, São Paulo 13083-970, Brasil.
[§]Departamento de Física, Universidade Federal de Viçosa, Viçosa, Minas Gerais 36570-900, Brasil.
[⊥]Centro de Pesquisas e Desenvolvimento Leopoldo Américo Miguez de Mello, Petrobras, Rio de Janeiro, Rio de Janeiro 21941-901, Brasil.
[¶]Instituto de Química de São Carlos, Universidade de São Paulo, São Carlos, São Paulo 13566-590, Brasil.
[∥]Instituto Nacional de Ciência e Tecnologia em Bioanalítica, Campinas, São Paulo 13083-970, Brasil.

**ABSTRACT:** We address a novel method for analytical determinations that combines simplicity, rapidity, low consumption of chemicals, and portability with high analytical performance taking into account parameters such as precision, linearity, robustness, and accuracy. This approach relies on the effect of the analyte content over the Gibbs free energy of dispersions, affecting the thermodynamic stabilization of emulsions or Winsor systems to form microemulsions (MEs). Such phenomenon was expressed by the minimum volume fraction of amphiphile required to form microemulsion ($\Phi_{ME}$), which was the analytical signal of the method. Thus, the measurements can be taken by visually monitoring the transition of the dispersions from cloudy to transparent during the microemulsification, like a titration. It bypasses the employment of electric energy. The performed studies were: phase behavior, droplet dimension by dynamic light scattering, analytical curve, and robustness tests. The reliability of the method was evaluated by determining water in ethanol fuels and monoethylene glycol in complex samples of liquefied natural gas. The dispersions were composed of water−chlorobenzene (water analysis) and water−oleic acid (monoethylene glycol analysis) with ethanol as the hydrotrope phase. The mean hydrodynamic diameter values for the nanostructures in the droplet-based water−chlorobenzene MEs were in the range of 1 to 11 nm. The procedures of microemulsification were conducted by adding ethanol to water−oleic acid (W−O) mixtures with the aid of micropipette and shaking. The $\Phi_{ME}$ measurements were performed in a thermostatic water bath at 23 °C by direct observation that is based on the visual analyses of the media. The experiments to determine water demonstrated that the analytical performance depends on the composition of ME. It shows flexibility in the developed method. The linear range was fairly broad with limits of linearity up to 70.00% water in ethanol. For monoethylene glycol in water, in turn, the linear range was observed throughout the volume fraction of analyte. The best limits of detection were 0.32% v/v water to ethanol and 0.30% v/v monoethylene glycol to water. Furthermore, the accuracy was highly satisfactory. The natural gas samples provided by the Petrobras exhibited color, particulate material, high ionic strength, and diverse compounds as metals, carboxylic acids, and anions. These samples had a conductivity of up to 2630 μS cm$^{-1}$; the conductivity of pure monoethylene glycol was only 0.30 μS cm$^{-1}$. Despite such downsides, the method allowed accurate measures bypassing steps such as extraction, preconcentration, and dilution of the sample. In addition, the levels of robustness were promising. This parameter was evaluated by investigating the effect of (i) deviations in volumetric preparation of the dispersions and (ii) changes in temperature over the analyte contents recorded by the method.

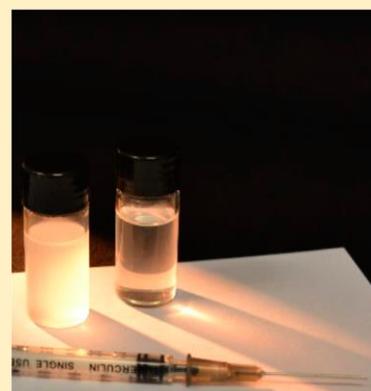

Analytical platforms for rapid tests are part of an important current research field that aims to perform in situ measurements, especially experiments such as urinalysis, food safety analysis, immunoassays, veterinary diagnostics, biothreats, drug abuse analysis, and environmental monitoring, in the developing world. Such technology is attractive because it is cheap, fast, portable, and simple, bypassing any qualified operators. Nonetheless, the rapid test devices usually show poor analytical performance and enable only screening analysis with the intent to determine unconformities rapidly.[1,2] Herein, we address a new method to conduct precisely preliminary analytical determinations that represents a promising alternative for the development of point-of-use technologies. This relies on the thermodynamic stabilization of dispersions to form microemulsion, and it includes requirements of an ideal rapid test technique by combining simplicity, rapidity, low consumption of chemicals, reduced cost, and portability with a strong analytical performance that takes into account parameters such as precision, linearity, robustness, and accuracy.[3]

Applications involving microemulsion (ME) were prompted by the oil crises in 1973 and 1979 aiming to improve the oil recovery in the rock pores.[1] Currently, MEs are applied in a broad range of fields. One example is their employment as reaction medium in the synthesis of diverse species as (i) nanostructures,[4−10] (ii) patterned polymer films,[11,12] (iii) molecular imprinted polymers,[13] (iv) copolymers,[14] and (v) 3D opals material.[15] Additionally, MEs are applied for (i) drug delivery,[16−18] (ii) catalysis,[19] (iii) extraction of metal,[20] and (iv) electrokinetic[21] and NMR chromatography,[22] where they act as background electrolytes.

The use of a microemulsification-based tool (denominated as MEC) to perform quantitative determinations is being described for the first time. Such a phenomenon is related to thermodynamic stabilization of dispersions, emulsion or Winsor system, to obtain microemulsion (ME). This effect was expressed by the minimum volume fraction of amphiphile (AP) required to form ME ($\Phi_{ME}$). It is the analytical signal in MEC, which can be measured simply by visual inspection (direct observation), bypassing the use of instrumental detection. This is possible because the dispersions change from cloudy to transparent during the microemulsification (Figure 1) in a similar way to titration.

MEC was initially applied to analyze ethanol fuel adulteration by water using direct observation with the naked eye to measure $\Phi_{ME}$. We employed dispersions composed of water (W), chlorobenzene (O phase), and ethanol (AP) that acted as hydrotrope. Next, the method was used to determine monoethylene glycol in liquefied natural gas samples. Herein, water (W), oleic acid (O), and ethanol (AP) made up the systems. Usually, the ME are stabilized by adding AP containing both the long (surfactant) and short (cosurfactant) nonpolar chains. The cosurfactants are commonly monohydric alcohols such as ethanol and propanol.[23] In some cases, small chain amphiphiles are enough for the microemulsification.[24] They are known as hydrotropes, and MEs are called the free-surfactant microemulsions.[25−28] Such dispersions present a phase and tension behavior similar to the surfactant-based MEs. It indicates that the hydrotropes partition between the phase in excess (where these APs are monomerically dissolved) and the W−O interfaces.[28] Here, AP is adsorbed as oriented monolayers favoring the microemulsification process by decreasing the Gibbs free energy as discussed below.

The following tests were included herein: (i) phase behavior, (ii) droplet dimensions, (iii) analytical curves, (iv) robustness, and applications in order to determine (v) water in ethanol fuel and (vi) monoethylene glycol in natural gas exploration samples. The level of robustness was assessed considering variations in the analyte content owing to deviations in the ME preparation and temperature.

■ **PRINCIPLE**

MEC relies on the effect of the analyte content over the Gibbs free energy ($G$) of dispersions by modifying the surface area ($\sigma$) and/or interfacial tension ($\gamma_i$) as follows:[23]

$$dG = \gamma_i d\sigma \quad (1)$$

Mathematically, this effect was expressed by the minimum volume fraction of AP necessary to form ME, $\Phi_{ME}$, which is the analytical signal of the method. For larger values of $\sigma$ and $\gamma_i$, a higher $\Phi_{ME}$ will be required. Variations in $\sigma$ are due to changes in composition of the W−O mixture. The $\gamma_i$, in turn, is modified when the analyte content interferes with the total excess interfacial concentration ($\Gamma$) according to the Gibbs adsorption equation:

$$d\gamma_i = -\sum_i \Gamma_i d\mu_i \quad (2)$$

where $\mu_i$ is the chemical potential of the species i. In this case, the analyte is usually adsorbed at the water−oil (W−O) interfaces to form microemulsion by reducing $\gamma_i$. This phenomenon is called the surface activity, and it decreases the interfacial tension because of the raise in surface pressure ($\pi$), according to[23]

$$\gamma_i = \gamma_o - \pi \quad (3)$$

where $\gamma_o$ is the initial interfacial tension before AP adsorption. The $\pi$ arises from lateral interactions between the polar ($\pi_w$) and nonpolar ($\pi_o$) groups of the amphiphile molecules so that

$$\pi = \pi_w + \pi_o. \quad (4)$$

Mixtures composed of W and O phases present interfacial tensions on the order of 30 to 50 mN m$^{-1}$ for nonionic APs. The thermodynamic stability of these W−O systems is ensured by reducing its interfacial tension to values of approximately 10$^{-3}$ to 10$^{-5}$ mN m$^{-1}$ through the addition of AP.[24] Lastly, it is relevant to highlight that the hydrotropes have lower surface activity and produce neither micelle nor lyotropic liquid crystal when compared to the surfactants.[28] More theoretical considerations about the subject in detail are available in the Supporting Information.

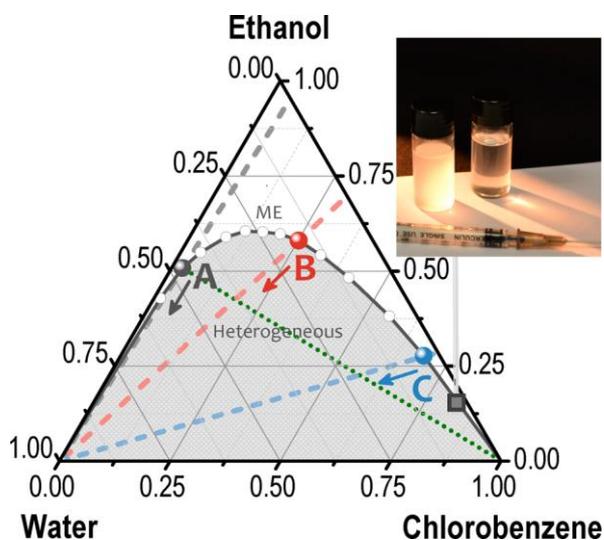

Figure 1. Phase behavior of the water−ethanol−chlorobenzene system at 23 °C. A, B, and C are the ME regions tested for quantitative analyses. Inset: photo illustrating the transition from cloudy (left, unstable dispersion) to transparency (right, ME) that occurs on the binodal curve. The dispersions for DLS measures were attained in the A and C regions along the green and blue dashed lines, respectively. Their compositions are placed in the white parts of the diagram (related to ME) which are highlighted in the inset of Figure 2. Additionally, the gray, red, and blue arrows indicate the displacement of the phase behavior with the addition of water in ethanol for the A, B, and C points of the binodal curve, respectively.

## ■ EXPERIMENTAL SECTION

**Chemicals.** Ethanol and silicon dioxide particles (for analyses by DLS) were purchased from Merck (Whitehouse Station, NJ) and Sigma-Aldrich Chemical Co. (St. Louis, MO), respectively. Chlorobenzene was acquired from Synth (Diadema, São Paulo, Brazil). Deionized water (Milli-Q, Millipore Corp., Bedford, MA) was attained with resistivity values higher than 18 MΩ cm.

**Dimensional Characterization of the Droplets in ME.** The diameter of the nanostructures in discontinuous ME applied to determine water in ethanol fuel samples was analyzed by a light scattering method using dynamic light scattering (DLS) appliances of the Brookhaven Instruments Corp. TurboCorr (NY, USA) and Malvern Zetasizer Nano-S (Worcestershire, UK). All of the samples were placed in 20 mL cylindrical glass cuvettes for Brookhaven setup and in glass square cuvettes for Zetasizer equipment. These samples were kept at 23.0 ± 0.5 °C and filtered utilizing 0.22 μm PVDF syringe filters. Due to the large vapor pressure of the organic compounds, special care was taken to prevent any loss of solvent. As a consequence, the filtration and addition of the liquids were made in a closed circuit with the aid of a pump. All the measures were performed in duplicate.

**Microemulsification.** The procedures of microemulsification were conducted by adding ethanol to W−O mixtures with the aid of micropipettes. First, specific volumes of W and O phases were transferred to glass bottles. Next, AP was added to each dispersion until the visual inspection of the cloudy-totransparent conversion. The required volume of AP phase (ethanol) for a given volume ratio of the W and O phases in this step can be represented as the volume fraction $\Phi_{ME}$, which was considered to be the analytical signal in MEC. The mixtures were vigorously shaken after adding the ethanol AP. As a consequence, the microemulsification was observed within a few seconds owing to convection mass transport generated by the shaking. The measurements of $\Phi_{ME}$ were carried out in a thermostatic water bath at 23 °C by direct observation that is based on the visual perception of the optical transition of the media from cloudy to transparent. All of the bottles were kept closed, but it is important to highlight that the solvent loss by evaporation did not represent a drawback due to the rapidity of the assays.

The analytical curve, application, and robustness relative to the determination of water in ethanol were investigated by utilizing dispersions prepared in three regions of the phase diagram (Figure 1). It was conducted to assess the analytical performance in each region, which was relative to water-rich (A) and oil-rich (C) domains and MEs with similar volumes of W and O (B). The W−O mixtures were prepared with 5.00% (A), 60.00% (B), and 95.00% v/v (C) oil to water ($\Phi_O$). In relation to the determination of monoethylene glycol, only region B was used. In this situation, conversely, the W−O mixtures were prepared with 50.00% v/v $\Phi_O$.

To obtain the analytical curves (relationship between $\Phi_{ME}$ and analyte volume fraction) in MEC, analyte standards with different concentrations are transferred to either W, O, or the AP phase depending on nature of the sample. Regarding the ethanol fuel, the sample is mainly composed of ethanol, thus acting as AP of the ME. Therefore, solutions of ethanol were prepared by dilution with different volume fractions of water ($\Phi_W$, analyte concentration) to get the analytical curves in the A, B, and C regions. Such solutions were transferred to the W− O mixtures to obtain the microemulsification. The linear range in the analytical signal allowed one to apply the method to real samples by using a titration-like experimental protocol. In this case, the sample was utilized directly as AP of the systems. It was added to W−O mixtures previously prepared until there was a cloudy-to-transparent change of the dispersions that corresponded to the microemulsification. The volume used in this step represents the analytical response in MEC, expressed as volume fraction $\Phi_{ME}$. For the measurements in natural gas, the sample is mainly composed of water and monoethylene glycol (it is not an AP). Hence, this acts as W phase of ME so that solutions of water were prepared by dilution with different volume fractions of monoethylene glycol ($\Phi_M$, analyte content). These solutions were transferred to O phase (50.00% v/v $\Phi_O$). After, pure ethanol was added to the W−O mixtures to obtain the microemulsification and, then, the analytical curve in region B. Lastly, the real samples were used as W of the dispersions.

**Analytical Curves.** The analytical calibration of the MEC was intended to calculate merit figures like correlation factor (R2), analytical sensitivity (S), and limit of detection (LOD), as well as to determine water in ethanol fuels and monoethylene glycol in natural gas exploration samples.

**Application to Determine Water in Ethanol Fuel.** MEC was applied to determine water in real and synthetic samples of ethanol fuel. Four solutions of ethanol were prepared in the laboratory by dilution with deionized water in volume fractions of 1.00%, 5.50%, 9.00%, and 17.00% v/v (synthetic). Three real samples, in turn, were purchased from different fuel stations. To assess the method accuracy, the water concentrations were also determined by Karl Fischer titration (Metrohm, Titrando 890, Herisau, Switzerland) which acted as the reference approach.[29] The statistical evaluation between the data obtained by both techniques was performed using Student's t tests at 95% confidence level. Lastly, the values of conductivity of the real samples were measured by using an AJ Micronal AJX-522 (São Paulo, Brazil) system.

**Application to Determine Monoethylene Glycol in Liquefied Natural Gas.** Tests with liquefied natural gas samples to determine monoethylene glycol were performed in four samples provided by Petrobras, a Brazilian multinational energy corporation. To study the accuracy, the analyte contents were determined by iodometry classic titration (protocol developed by the Petrobras; its steps are confidential). Student's t tests at 95% confidence level were used again to assess the statistical comparison between the results. In addition, the samples were characterized in relation to the presence of anions, carboxylic acids, heavy metals, and silicon as discussed in the Supporting Information. Finally, the conductivities of these samples were measured through the AJ Micronal AJX-522 device.

**Robustness Tests.** The robustness of MEC was assessed by investigating the effect of a small change in the W−O mixture ratio and temperature. The level of robustness was expressed in terms of absolute error determined for $\Phi_W$ and $\Phi_M$ ($\Delta\Phi$, %v/v). Such errors were caused by modifications in $\Phi_O$ and temperature of ME. To evaluate the effect of the procedure in preparing W−O mixtures, $\Delta\Phi$ was calculated in relation to 5.00% v/v $\Phi_W/\Phi_M$ (reference value) considering relative standard deviations (RSD) of 5.00% and 10.00% v/v (in three different regions with $\Phi_O$ of 5.00% in A, 60.00% in B, and 95.00% v/v in C for the applications in ethanol and 50.00% v/v only for applications in natural gas samples). The $\Phi_W$ and $\Phi_M$ were experimentally determined from analytical curves. To investigate the temperature-function robustness, analytical curves were initially obtained at several temperatures. The used procedure was the same of that aforesaid at 23 °C. The $\Delta\Phi$ was related to 5.50%, 9.00%, and 14.00% v/v $\Phi_W$ at 23 °C for the water-applied system and, finally, to 30.00%, 60.00%, and 90.00% v/v $\Phi_M$ at 23 °C for the monoethylene glycolapplied system (reference values).

## ■ RESULTS AND DISCUSSION

**Application to Determine Water in Ethanol Fuel.**
*Phase Behavior.* The ternary phase diagram of the dispersions composed of water−ethanol−chlorobenzene is shown in Figure 1. The mixtures were prepared from different values of $\Phi_O$. A blank region from the water- to oil-rich domains can be observed. Such dispersions above the binodal curve are related to ME. The shaded regions, in turn, represent unstable dispersions. The average values of $\Phi_{ME}$ ($n$ = 4) were employed to plot the diagram with confidence intervals ($\alpha$ = 0.05) ranging from 0.10% to 0.22% v/v. MEs were prepared with volumes to have a final mixture with approximately 1.0 mL. The diagram in Figure 1 shows a typical phase behavior,[24] with lower efficiency (this decreases with $\Phi_{ME}$) in the region of bicontinuous MEs (B) which contain continuously interpenetrating domains (water and oil).

*Characterization of the Droplet Dimensions in ME.* Dispersions were prepared in the A and C regions along the green and blue dashed lines illustrated in Figure 1, respectively. Their compositions are placed in the white parts of the ternary phase diagram (relative to ME) which are highlighted in Figure 2. Briefly, solutions of 52.50% v/v ethanol to water (A) and 29.50% v/v ethanol to chlorobenzene (C) were prepared. Afterward, small amounts of the inner phase (oil for A and water for C) were added to generate ME.

The mean hydrodynamic diameters ($d_h$) for the nanostructures in the droplet ME are presented in Figure 2. Such values are functions of the volume fractions of oil ($\Phi_{O,D}$) and water ($\Phi_{W,D}$) to total dispersion. The diameters were in the range from 1 to 11 nm. As expected, their values increased with the fraction of added inner phase. The $d_h$ data reported herein are in agreement with those obtained by DLS[27,30,31] and fluorescence[32] for microemulsions based on hydrotropes (2−30 nm). Recently, the presence of structured aggregates of ethanol in octanol-rich domains was demonstrated by static light scattering (SLS) and DLS.[31] Two years later, such a system was characterized by molecular dynamics simulations.[33] Herein, aggregates with an emerging interface were observed.

More recently, images by transmission electron microscopy revealed dh values of up to 50 and 40 nm for the dispersions of water−ethanol−benzene[34] and water−n-propanol−oleic acid,[28] respectively. The $d_h$ achieved for A and C was empirically fitted to a shifted power-law function as shown in the Supporting Information. The diameter values that diverged in relation to this function were defined as the boundaries between the droplet and bicontinuous nanostructures. Such boundaries for the O−W-to-bicontinuous and W−O-to-bicontinuous transitions were estimated as 3.00% v/v $\Phi_{O,D}$ and 4.30% v/v $\Phi_{W,D}$, respectively.

*Analytical Curves.* The resulting analytical curves are depicted in Figure 3. The observed positive angular coefficients are due to the decrease in ethanol concentration as the $\Phi_W$ increases, requiring larger values of $\Phi_{ME}$ to stabilize the dispersions. The confidence intervals ($\alpha$ = 0.05, $n$ = 4) were around 0.20% v/v $\Phi_{ME}$ at all of the tested points.

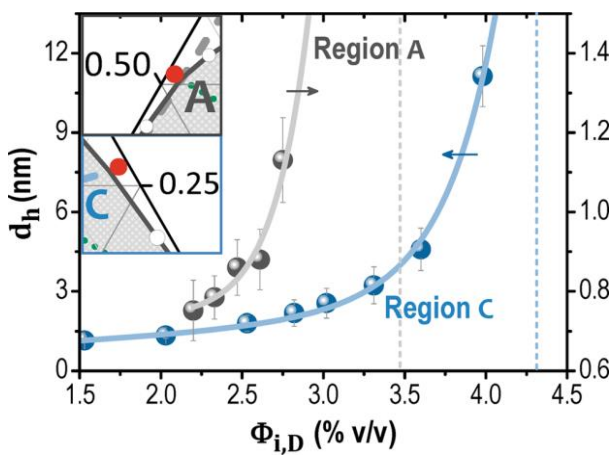

**Figure 2.** Values of $d_h$ for MEs in the A (gray data, right axis) and C (blue data, left axis) regions as functions of $\Phi_{i,D}$ ($\Phi_{O,D}$ for A and $\Phi_{W,D}$ for C). Insets, amplified phase diagram regions (highlighted by red circles) of the compositions tested by DLS. The dashed lines (gray for A and blue for C) are $\Phi_{i,D}$ values related to transitions for B structures.

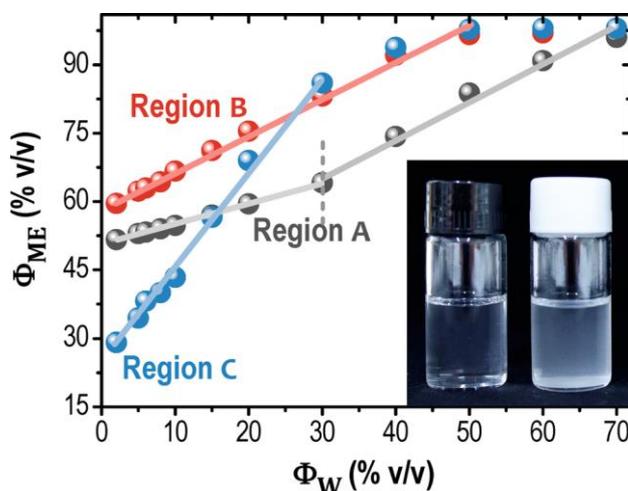

**Figure 3.** Analytical curves for A, B, and C regions. Inset: photo of the transition from transparency to cloudy for dispersions in A with 5.00% (left, ME) and 6.00% v/v $\Phi_W$ (right, unstable system). In both the cases, the ethanol volume needed to stabilize only emulsions with 5.00% v/v $\Phi_W$ was added. $R_2$: 0.9983 and 0.9915 (first and second linear range in A, respectively), 0.9973 (B), and 0.9956 (region C).

MEC had different levels of linearity, sensitivity, and detectability for the A, B, and C regions. The linear range was fairly broad with limits of linearity (LOLs) of 50.00% for B and 30.00% v/v for C. For A, two linear ranges were obtained with LOL values of 30.00% and 70.00% v/v.

The analytical sensitivities were 0.44 (first linear range for A), 0.80 (second linear range for A and B), and 2.04 (for C). The limits of detection (LOD) were calculated using two methods: (i) visual and (ii) signal/noise ratio. For the visual approach, $\Phi_{ME}$ fractions were added in compositions containing 5.00% v/v $\Phi_W$ (40.00% v/v for the second linear range in A). In these cases, MEs formed. Then, the ethanol was diluted using volume fractions of water above 5.00% and 40.00% v/v. Next, the same values of $\Phi_{ME}$ adopted in the previous step were added to A, B, and C. Here, the stabilization of the medium is not expected because the reduction in the ethanol content as $\Phi_W$ is raised. Consequently, the dispersions must remain heterogeneous. The minimum changes in $\Phi_W$ that allowed us to visualize these cloudy heterogeneous media were defined as LOD according to the visual method. Their values were 1.00% (first linear range for A), 0.60% (second linear range for A), 0.60% (B), and 0.40% v/v (C). In Figure 3, a photo displays the difference in the visual aspect between a microemulsion (5.00%) and a heterogeneous dispersion (6.00% v/v $\Phi_W$). It is associated with the LOD condition taking into account the first linear range in A. Finally, the LODs obtained from the signal/noise ratio were equal to 1.41% (A), 0.73% (B), and 0.32% v/v. The values of the $\Phi_{ME}$ accomplished when considering pure ethanol as AP ($\Phi_W$ = 0.00% v/v, $n$ = 4) were adopted to calculate the blank standard deviations. For the different sensitivities and detectabilities in the investigated regions, our hypothesis relates to the effect of the AP-added water over the microemulsification process by changing the surface area and pressure. It is cited in more detail in the Supporting Information.

*Applications.* The resulting data for the samples of ethanol fuel are demonstrated in Table 1. The results attained by Karl Fischer and MEC were consistente with the water concentrations prepared in the laboratory by diluting the ethanol AP. As a consequence, there was no statistically meaningful difference among the analyte contents in the synthetic samples and those determined by both tested methods. Regarding the real samples, the data obtained by MEC in the B and C regions were in agreement with those determined by Karl Fischer. The results in A, in turn, were not consistent with those obtained by this reference method. Considering the excess of water in A, such discrepancy was likely due to the ionic strength of the ethanol fuel samples. Investigations disclosed the presence of diverse ions in these samples, including $NO_3^-$, $K^+$, $Ca^{2+}$ (0.49 to 3.51 mg $L^{-1}$),[35] $Cu^{2+}$, $Zn^{2+}$, $Ni^{2+}$, and $Fe^{3+}$ (8 to 57 mg $L^{-1}$).[36] The real samples had an average conductivity of 1.28 μS $cm^{-1}$, whereas it was measured as only 0.37 μS $cm^{-1}$ for pure ethanol.

**Table 1. Results of the Determination of $H_2O$ in Ethanol ($\Phi_W$, % v/v) Using Karl Fischer Titration ($n$ = 3) and MEC ($n$ = 4)[a]**

| samples | Karl Fischer (%, v/v) (±0.02) | MEC (%, v/v) A (±0.07) | B (±0.05) | C (±0.08) |
|---|---|---|---|---|
| $S_1$ | 5.18 | 5.20 | 5.33 | 5.31 |
| $S_2$ | 8.51 | 8.43 | 8.54 | 8.52 |
| $S_3$ | 17.24 | 17.44 | 16.98 | 17.94 |
| $S_4$ | 45.73 | 45.83 | 45.63 | 45.41 |
| $R_1$ | 5.21 | 36.07 | 5.25 | 5.00 |
| $R_2$ | 5.01 | 40.37 | 5.17 | 4.90 |
| $R_3$ | 4.86 | 11.13 | 4.86 | 4.92 |

[a]The confidence intervals were calculated for α = 0.05. Synthetic samples: $S_1$, 5.50%; $S_2$, 9.00%; $S_3$, 17.00%; and $S_4$, 45.50% v/v. Real samples of ethanol fuel: $R_1$–$R_3$.

Determination of water in ethanol standard samples was cited in the literature using different methods, such as (i) enthalpimetry,[37] (ii) evanescent field absorption spectroscopy,[38] (iii) spectroscopy in the infrared employing pattern recognition techniques,[39] (iv) capillary gas chromatography based on ionic liquid with a thermal conductivity detector,[40] and (v) a photothermal transducer.[41] The gas chromatography reached a LOD of 880 ppm water in ethanol.[40] Recently, Ribeiro et al. reported a protocol to determine the adulteration of ethanol fuel by water.[29] It relies on solubility differences of sodium chloride in ethanol and water. The concentrations of this salt in saturated media were monitored by conductivity. LOL and LOD were 16.00% and 0.05% v/v, respectively.

*Robustness Tests.* Concerning the procedure of W–O mixture preparation, the ΔΦ values were 1.81% (A), 0.59% (B), and 0.93% v/v (C) considering 10.00% v/v RSD. Such errors for 5.00% v/v RSD, in turn, were 0.81% (A), 0.29% (B), and 1.28% v/v (region C). Such differences are due to the effects of $\Phi_O$ over $\Phi_{ME}$ (see Figure S3 in the Supporting Information). In region B, e.g., $\Phi_{ME}$ is practically invariable with the values of $\Phi_O$ explaining the low RSD obtained in this case. In relation to the investigation about the temperature-function robustness, the resulting analytical curves are depicted in Figure 4. The obtained ΔΦ values for the changes of 23 to 20 °C are shown in the inset of Figure 4, whereas those related to the heating of 23 to 26 °C are in the Supporting Information. In general, the results in region B and, especially, region C exhibited the best robustness owing to their greater sensitivities (its values are cited in the Figure 4 caption). ΔΦ ranged from only 0.30% to 1.10% v/v in C. Despite the satisfactory data, high-precision analyses may require the use of analytical curves for certain differences in temperature.

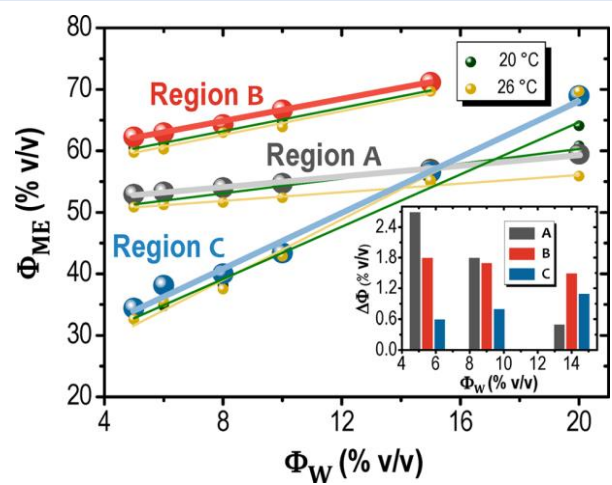

**Figure 4.** Analytical curves at different temperatures to investigate the robustness level. Inset: values of ΔΦ as a function of $\Phi_W$ in three regions of the analytical curves (5.50%, 9.00%, and 14.00% v/v $\Phi_W$) for the change of 23 to 20 °C in the A, B, and C regions. The data in gray (A), red (B), and blue (region C) were accomplished at 23 °C. Those in green and yellow are relative to 20 and 26 °C, respectively. All of the $R_2$ values were larger than 0.9900. The values of S, in turn, were: (i) 0.60 (20 °C), 0.43 (23 °C), and 0.36 (26 °C) for A, (ii) 0.94 (20 °C), 0.90 (23 °C), and 0.96 (26 °C) for B, and (iii) 2.12 (20 °C), 2.27 (23 °C), and 2.47 (26 °C) for region C. The ΔΦ parameter is given in the module; some of its obtained values were negative.

**Application to Determine Monoethylene Glycol in Liquefied Natural Gas.** The application of MEC to determine monoethylene glycol in liquefied natural gas exploration samples was intended to test the potential of the developed method for complex samples. As shown below, the samples provided by Petrobras had color, particulate material, high ionic strength, and diverse compounds such as metals, carboxylic acids, and anions. Despite these downsides, MEC allowed accurate measurements bypassing steps such as extraction, preconcentration, and dilution of the sample; the $\Phi_M$ ranged from approximately 57% to 96% v/v.

The monoethylene glycol is applied to the production lines of natural gas to prevent the clogging of pipes due to hydrate formation. Conversely, the monoethylene glycol acts as an undesirable contaminant causing loss of quality of the final product and corrosion in the piping system and poisoning the catalyst. Hence, this dialcohol is removed from the natural gas production lines after its use. The monitoring of such compound is, then, necessary in order to test the effectiveness of its regeneration procedure as well as to ensure a final product that complies with the quality standards required by the industry and consumer market.

*Phase Behavior.* Phase diagram of the water−ethanol−oleic acid dispersions is presented in Figure 5. The microemulsification was repeated by three times in each point. The $\Phi_{ME}$ average values were used to plot the diagram with confidence intervals ($\alpha$ = 0.05 and $n$ = 4) between 0.13% and 0.21% v/v $\Phi_{ME}$. Volumes of approximately 1.0 mL were used to prepare $\Phi_{ME}$, and approximately 1.0 mL was used to prepare the dispersions. The binodal curve did not have a typical profile for neutral AP, with greater efficiency in region A. Here, an essential modification was the substitution of chlorobenzene (used for water analysis) by oleic acid as O phase. It reduces risks related to human health and the environment.

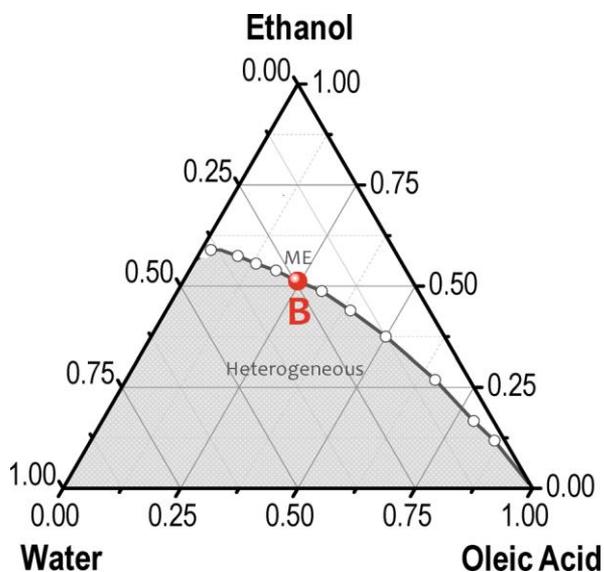

**Figure 5.** Phase diagram of the water−ethanol−oleic acid system at 23 °C. Only the region B was employed for the studies of analytical curve, application, and robustness related to the determination of monoethylene glycol.

*Analytical Curve.* The obtained analytical curve is depicted in Figure 6. Herein, the negative angular coefficients are likely owing to the decrease in $\gamma_i$ value with the successive additions of monoethylene glycol in water (W phase). The confidence intervals ($\alpha$ = 0.05, $n$ = 4) were approximately 0.17% v/v $\Phi_{ME}$ at all of the points. The linear range extended throughout the volume fraction of analyte, with two regions of distinct analytical sensitivities. S values were 0.20 (0.00% to 60.00%) and 0.40 (60.00% to 100.00% v/v $\Phi_M$). LODs were calculated again using the two methods: (i) visual and (ii) signal/noise ratio. LODs according to the visual method were 1.20% (first linear range) and 0.60% (second linear range). The limit of detection obtained from the signal/noise ratio, in turn, was 0.31% v/v. The $\Phi_{ME}$ values relative to the use of pure water as W ($\Phi_M$ = 0.00% v/v, $n$ = 4) were employed to calculate the blank standard deviations. Concerning the different sensitivity that was verified in region B of the diagram, the effect of the water on microemulsification process is presumably the reason again for such a phenomenon as described in the Supporting Information.

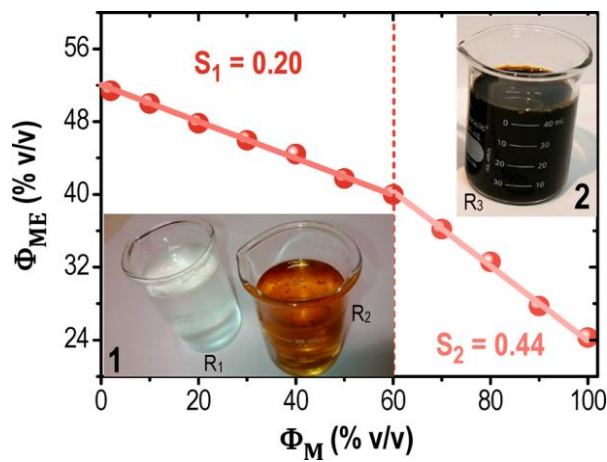

**Figure 6.** Analytical curve for water−ethanol−oleic acid ME. Insets: 1 and 2, photos of the real samples provided by Petrobras (R$_4$ is visually similar to R$_1$). $R^2$: 0.9969 for first and 0.9965 for second linear range. Their S values (S$_1$ and S$_2$) are depicted.

*Application.* The resulting data for the natural gas samples are shown in Table 2. There was no statistical difference between the analyte concentrations (%, m/V) achieved by iodometry and MEC. It is important to highlight that the method allowed one to perform experiments without steps such as extraction, dilution, and preconcentration of the samples and obtain accurate results. As cited above, these samples presented color and particulate changes as illustrated in Figure 6.

**Table 2. Determination of Monoethylene Glycol ($\Phi_M$, % m/v) in Liquefied Natural Gas Samples (R1−R4) Using Iodometry Titration ($n$ = 3) and MEC ($n$ = 4) in Region B**[a]

| samples | iodometry (%, m/v) (±1.0) | MEC (%, m/v) (±1.5) | $\kappa$ ($\mu$S cm$^{-1}$) |
|---|---|---|---|
| R1 | 96.2 | 97.7 | 54.1 |
| R2 | 60.7 | 59.6 | 2630.0 |
| R3 | 57.3 | 57.8 | 745.0 |
| R4 | 54.7 | 55.7 | 349.0 |

[a]The confidence intervals were calculated for $\alpha$ = 0.05.

We conducted direct analyses with exception of one sample (Figure 6, inset 2). Herein, a simple filtration was made using 0.20 μm PTFE syringe filters. Furthermore, the samples had high ionic strength and diverse compounds. The conductivity ($\kappa$) values are presented in Table 2. Lastly, assays made by Petrobras revealed the presence of metals ($Na^+$, $K^+$, $Mg^{2+}$, $Ca^{2+}$, $Ba^{2+}$, $Sr^{2+}$, and $Fe^{3+}$), anions ($Cl^-$, $Br^-$, and $SO_4^{2-}$), carboxylic acids (glycolate, formate, acetate, propionate, and butyrate), and silicon in the samples (Supporting Information).

*Robustness Tests.* In relation to the procedure of W−O mixture preparation, the values of $\Delta\Phi$ were only 0.44% and 2.39% v/v for RSD of 5.00% and 10.00% v/v, respectively. Regarding the tests with different temperatures (19 to 29 °C), the obtained analytical curves are presented in Figure 7.

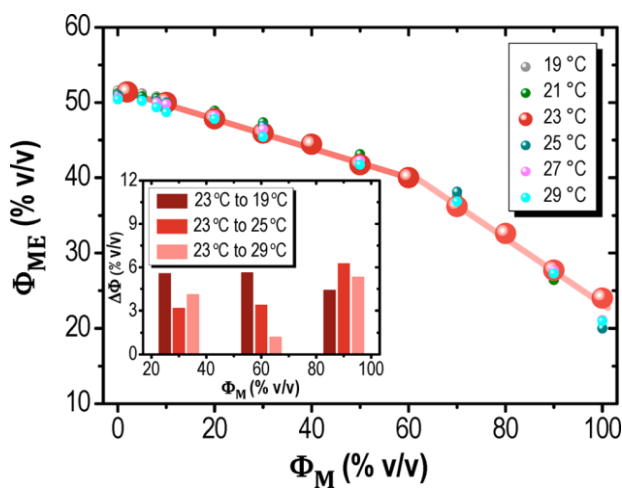

**Figure 7.** Analytical curves at different temperatures. Inset: values of $\Delta\Phi$ as a function of $\Phi_M$ for 30.00%, 60.00%, and 90.00% v/v $\Phi_M$ to different changes in temperature. All the values of $R_2$ were larger than 0.9900. In the inset, the regression straight equation of the second linear range was used for 60.00% v/v $\Phi_M$ because of its greater S. $\Delta\Phi$ is presented in the module; some values were negative.

The $\Delta\Phi$ values for the modifications of 23 °C to 19 °C, 23 °C to 26 °C, and 23 °C to 29 °C are depicted in the Figure 7 inset. $\Delta\Phi$ changed from 1.24% to 6.30% v/v. Such results demonstrate that the method is potentially robust in addition to being simple and rapid.

## ■ CONCLUSIONS

A new strategy for quantitative determinations is reported in this Article. The MEC represents an exponential contribution for the development of rapid test platforms taking into account its strong analytical performance, rapidity, and simplicity and the fact that no sophisticated instrumentation or equipment is required. For rapid test measures, another important feature is the compatibility of the approach with small volumes of sample. Tests performed with 500 μL Eppendorf tubes confirmed that volumes on the order of 20 μL for total dispersion enable direct observations still for the measurement of $\Phi_{ME}$. Despite its simplicity, the MEC provided precise, robust, and accurate data with satisfactory detectability and sensitivity for determination of water in ethanol fuel and monoethylene glycol in complex samples of liquefied natural gas.

The robustness in MEC depends on parameters such as the precision of the preparation of the dispersions, temperature, and ionic strength. Data obtained in this paper for oil-rich (C) domains and MEs containing similar volumes of W and O (region B) indicated satisfactory robustness levels with respect to deviations in the dispersion preparation and temperature. Concerning the ionic strength effect, our data were also promising. For ethanol fuels (average conductivity of 1.28 μS cm$^{-1}$), the results in the B and C regions presented a good accuracy. The accuracy of MEC was impressively proven to be great in the region B for determination of monoethylene glycol in natural gas samples. These exhibited a conductivity of up to 2630 μS cm$^{-1}$; the conductivity of pure monoethylene glycol was only 0.30 μS cm$^{-1}$. Furthermore, the signals were broadly linear with LOL of up to 70.00% v/v water to ethanol. For monoethylene glycol in water, in turn, the linear range was observed throughout the volume fraction of analyte.

MEC also permits screening analysis. With the intent to rapidly detect unconformities, an identical procedure related to that employed for the visual method-based LOD calculation should be applied. In Figure 2, e.g., it is possible to state that the second sample in the inset (right) has a water content higher than 5.00% v/v. A single sample pipetting is necessary for screening analyses. Since the W−O mixture could be previously prepared, the assay is simple and rapid. For precise determinations, in turn, a titration-like step is required to obtain the minimum volume fraction of AP needed to generate ME. Herein, in situ analyses could be easily performed with the aid of digital micropipettes (it includes a digital dial micrometer) or microfluidics (integrating optical detection to measure ME). With the conversion of the bulk-based analyses in microfluidics, further advantages include faster analyses, smaller consumption of chemicals (femtoliter to nanoliter), and an improvement in the analytical performance.[42] A downside of this system is the loss of sensitivity by the radiation scattering in chip. One alternative to reduce this phenomenon is the employment of optical fibers to guide the excitation light and collect the emitted radiation.[43] In addition, the laminar flow attained in microchannels creates a diffusion-limited mixing. Consequently, micro mixers should surely be integrated in the microdevice for the rapid homogenization of the dispersions, as well as passive approaches such as lamination and chaotic advection.[42]

All of the aforementioned features contribute to simplicity and rapidity of the developed method. In addition, depending on the nature of the sample, the analyte can be added either in the W, O, or AP phase. When the sample acts like a W phase, e.g., analyte standards are transferred to W to get the analytical curve. For application to synthetic and real samples, these are then mixed to the O phase before the addition of pure amphiphile. Herein, the sample is the W phase of the microemulsion, as was the case of the natural gas samples. The approach is, hence, promising to analyze polar, nonpolar, and amphiphilic matrices. The diverse analytical performances for the different compositions are other aspects that demonstrate flexibility to the MEC. When the quantity of available sample is very small, e.g., a composition of the ME that requires less analyte volume should be adopted.

This Article raises questions which encourage new scientific investigations, such as more detailed studies about the effect of diverse factors over the analytical performance. In addition, other types of AP could be investigated to improve the analytical performance, including cationic, anionic, amphoteric, and biodegradable APs, hydrotropes and surfactants.

About its application, the MEC is promising even considering the measurements in a biological medium. In principle, the analytical responses for the detection of biomolecules in complex matrices will not be specific because all of the compounds will change the dispersion $G$. Herein, an innovative alternative is the use of APs that can also act as analyte chemical receptors. Therefore, the changes in $\gamma_i$ will be selective to the biomolecular interactions. We believe such an approach will enable the employment of the MEC in complex matrices. Such a method could be an effective potential alternative for the development of point-of-care testing technologies which represent 36% of the global in vitro diagnostics market. In 2011, this market was valued at $44 billion. Rapid test devices are attractive because they permit low cost diagnostics in the developing world.[1] Conversely, only a few methods have been implemented with commercial success. This is owing to market barriers such as the production cost of the analytical devices, inhibiting the creation of profitable businesses.[44]

## ■ ASSOCIATED CONTENT

*S Supporting Information

Additional information as noted in text. This material is available free of charge via the Internet at http://arXiv.org.

## ■ AUTHOR INFORMATION

**Corresponding Author**
*E-mail: kubota@iqm.unicamp.br.
**Notes**
The authors declare no competing financial interest.

## ■ ACKNOWLEDGMENTS

Financial support for this project was provided by Petrobras (Grant Nr. 2012/00029-6), Financiadora de Estudos e Projetos (FINEP, Grand PROINFRA 01/2009), and Conselho Nacional de Desenvolvimento Cientifí co e Tecnológico (CNPq). Professor Dr. Edvaldo Sabadini from Universidade Estadual de Campinas is also thanked for his help with key theoretical discussions.

## ■ NOMENCLATURE

| | |
|---|---|
| $\gamma_i$ | interfacial tension. |
| $\pi$ | surface pressure. |
| $\Phi_{ME}$ | minimum volume fraction of amphiphile needed to generate ME. |
| $\Phi_O$ | volume fraction of oil to W phase. |
| $\Phi_W$ | volume fraction of water to ethanol. |
| $\Phi_M$ | volume fraction of monoethylene glycol to water. |
| $\Phi_{W,O}$ | volume fraction of water to O phase. |
| $S$ | analytical sensitivity. |
| LOD | limit of detection. |

# Microemulsification: a novel approach for analytical determinations

Renato Sousa Lima,[†,‡,£] Leandro Yoshio Shiroma,[†,‡] Alvaro Vianna Novaes de Carvalho Teixeira,[§] José Roberto de Toledo,[§] Bruno Charles do Couto,[φ] Rogerio Mesquita de Carvalho,[φ] Emanuel Carrilho,[£,\\] Lauro Tatsuo Kubota*[‡,£] and Angelo Luiz Gobbi[†]

[†] Laboratorio de Microfabricacao, Laboratorio Nacional de Nanotecnologia, Centro Nacional de Pesquisa em Energia e Materiais, Campinas, SP, Brasil.
[‡] Instituto de Quimica, Universidade Estadual de Campinas, Campinas, SP, Brasil.
[§] Departamento de Fisica, Universidade Federal de Vicosa, Vicosa, MG, Brasil.
[φ] Centro de Pesquisas e Desenvolvimento Leopoldo Americo Miguez de Mello, Petrobras, Rio de Janeiro, RJ, Brasil.
[\\] Instituto de Quimica de Sao Carlos, Universidade de Sao Paulo, Campinas, SP, Brasil.
[£] Instituto Nacional de Ciencia e Tecnologia em Bioanalitica, Campinas, SP, Brasil.
* Tel: +55 (19) 3521-3127. E-mail: kubota@iqm.unicamp.br.

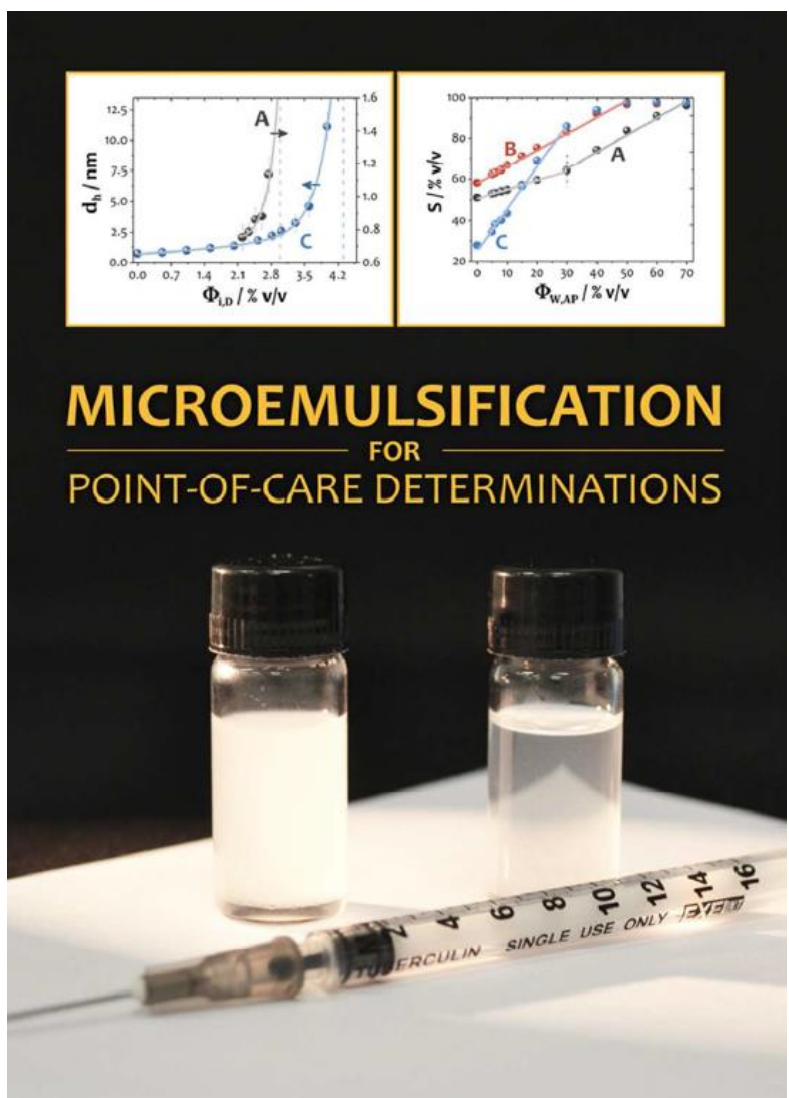

TABLE OF CONTENTS

■ **Theory**

ME are thermodynamically stable dispersions composed of at least a hydrophilic and a hydrophobic species, usually called the water (W) and oil (O) phases, respectively. This dispersion is stabilized by amphiphilic compounds that are adsorbed at the W-O interfaces, generating oriented monolayers. ME exhibit different features with respect to emulsions: (i) their particles in droplet-based systems present nanometer dimensions with radii of 1 to 300 nm, whereas these values range from 1 to 10 μm in the emulsions; (ii) ME and emulsions are optically transparent and cloudy, respectively; and (iii) only the microemulsion systems are thermodynamically stable because of the higher surface activity.[1] The following were included in this section: (i) quantitative (Gibbs adsorption equation) and (ii) qualitative (surface pressure) reasons for reduction of the interfacial tension with the amphiphile content and (iii) thermodynamic stabilization of heterogeneous dispersions generating ME (microemulsification).

**1. Reduction of the interfacial tension with the amphiphile content: Gibbs adsorption equation**

The Gibbs adsorption equation shows the mathematical relationship of the interfacial tension ($\gamma_i$) with the amphiphile content. This is achieved from the fundamental equation of thermodynamics which defines the Gibbs free energy (*G*) controlling the system equilibrium state functions: pressure (*p*), temperature (*T*), matter quantity, and interfacial area ($\sigma$).[1-3] The derivation of this fundamental equation will be discussed.

*Variation of the Gibbs free energy with pressure and temperature*

Unlike the entropy, Gibbs free energy expresses the spontaneity of the phenomena based on properties of the system only. Mathematically, taking into account infinitesimal changes of *G* keeping pressure and temperature constants, we have:[2]

$$dG = dH - TdS \quad (1)$$

where $dH$ and $dS$ are infinitesimal changes of enthalpy and entropy of the system, respectively. The equation for $dH$ is given by:

$$dH = dU + p\,dV \quad (2)$$

being *dU* the infinitesimal change of internal energy of the system added to the product of pressure and variation of volume. Thus:

$$dG = dU + p\,dV - TdS \quad (3)$$

Considering infinitesimal modifications in *p* and *T*, the derivation of (3) yields:

$$dG = dU + p\,dV + V\,dp - TdS - S\,dT \quad (4)$$

*dU* is defined as the system total energy. Its relationship with the work ($w$) and heat ($q$) is:

$$dU = dw + dq + dw_e \quad (5)$$

so that $w$ and $w_e$ are called as expansion (there is volume change) and extra (there is no volume change) work, respectively.

Thermodynamically reversible transformations can be inverted from infinitesimal changes of the system state functions. Then, such functions (*p* and *q*) are similar in relation to those of the neighborhood ($p_{ext}$ and $q_{ext}$). Thus, from fundamental concepts of thermodynamics, *dw* and *dq* can be calculated as:

$$dw = -p_{ext}dV = -p\,dV \quad (6)$$

and

$$dS = \frac{dq_{rev}}{T} = \frac{dq}{T} \quad (7)$$

so that:

$$dq = TdS \quad (8)$$

Substituting (6) and (8) in (5), we have for reversible phenomena with absence of $w_e$:

$$dU = -pdV + TdS \quad (9)$$

Substituting (9) in (4):

$$dG = -p\,dV + TdS + pdV + Vdp - TdS - SdT \quad (10)$$

with:

$$dG = V\,dp - S\,dT. \quad (11)$$

*Variation of the Gibbs free energy with amphiphile content*

The chemical potential of the species i ($\mu_i$) represents its capacity to change extensive parameters such as *G* at specific conditions. More formally, $\mu_i$ represents the variation of the Gibbs free energy for the addition of 1 mol of the species i to the system considering *p*, *T*, and *n'* (quantity of matter for all other components) constants. Mathematically, $\mu_i$ is the angular coefficient of *G* as a function of the number of moles of the added species i ($\mu_i$). Thus:[2]

$$\mu_i = \left(\frac{\partial G}{\partial n_i}\right)_{p,T,n'}. \quad (12)$$

Taking into account infinitesimal changes of $n_i$:

$$dG = \sum_i \mu_i dn_i. \tag{13}$$

*Variation of Gibbs free energy with interface área*

For interfaces, the work necessary to increase a surface area in reversible process at constant temperature ($w_e$) is given by:[3]

$$dw_e = \gamma_i d\sigma. \tag{14}$$

Considering a reversible process with *p* and *T* constant. We will have after including (6), (8), and (14) in (5):

$$dU = -pdV + TdS + \gamma_i d\sigma. \tag{15}$$

Substituting (15) in (3):

$$dG = -pdV + TdS + \gamma_i d\sigma + pdV - TdS. \tag{16}$$

then:

$$dG = \gamma_i d\sigma. \tag{17}$$

*Fundamental equation of thermodynamics*

Considering the variations of the Gibbs free energy with pressure, temperature, amphiphile content, and interface area as aforementioned, the fundamental equation of thermodynamics is obtained as highlighted in (18).[3]

$$dG = Vdp - SdT + \sum_i \mu_i dn_i + \gamma_i d\sigma. \tag{18}$$

*Gibbs adsorption equation*

For infinitesimal changes of $\mu_i$ and $\gamma_i$:

$$dG = Vdp - SdT + \sum_i \mu_i dn_i + \sum_i n_i d\mu_i + \gamma_i d\sigma + \sigma d\gamma_i. \tag{19}$$

Taking into account dispersions in the equilibrium (reversible process) to *p* and *T* constants with fixed values of $n_i$ and $\sigma$, we have:

$$\sum_i n_i d\mu_i + \sigma d\gamma_i = 0. \tag{20}$$

and finally:

$$d\gamma_i = -\sum_i \frac{n_i}{\sigma} d\mu_i \tag{21}$$

or

$$d\gamma_i = -\sum_i \Gamma_i d\mu_i \tag{22}$$

where $\Gamma_i$ is the excess interfacial concentration of the species i. This is given by:

$$\Gamma_i = \frac{n_i}{\sigma}. \tag{23}$$

In dispersions composed of water (W) and oil (O), $\Gamma_i$ corresponds to the quantity of i that exists in excess in the W-O interface when compared to that interface if it were mathematically plane. Actually, the region between the phases W and O has variable composition and thickness depending on factors such as dimensions and the nature of the components presente in the system.

The equation (22) is the generic approach of the Gibbs adsorption equation. Continuing, for three-components microemulsions (W, O, and amphiphile, AP):[3]

$$d\gamma_i = -\Gamma_W d\mu_W - \Gamma_O d\mu_O - \Gamma_{AP} d\mu_{AP} \qquad (24)$$

with $\Gamma_W = \Gamma_O = 0$. Thereby:

$$d\gamma_i = -\Gamma_{AP} d\mu_{AP} \qquad (25)$$

in which we observe the linear reduction of the interfacial tension with the amphiphile concentration. Such linear relationship is valid for a broad range of AP contents. In concentrations rather low (the most of the AP is monomerically solubilized; mass fraction values on the order of $10^{-5}$ for nonionic surfactants) and excessive (saturation of the surfactants in both the solvente and interface W-O generating micelles; this is not observed in hydrotrope-based dispersions), (25) is not correct due to the monomeric solubility and the formation of micelles, respectively.[1]

Continuing, the chemical potential of the AP phase is defined like:

$$d\mu_{AP} = RT\, d\ln(a_{AP}) \qquad (26)$$

where $R$ is the universal gas constant and $a_{AP}$ is the chemical activity of the amphiphile adsorbed in the W-O interface. Replacing (26) in (25):

$$d\gamma_i = -\Gamma_{AP} RT\, d\ln(a_{AP}) \qquad (27)$$

with:

$$\Gamma_{AP} = -\frac{1}{RT}\frac{d\gamma_i}{d\ln(a_{AP})} \qquad (28)$$

and:

$$\Gamma_{AP} = -\frac{a_{AP}}{RT}\frac{d\gamma_i}{da_{AP}} \qquad (29)$$

For diluted solutions where chemical activity and concentration (*c*) are similar, we have at last:

$$\Gamma_{AP} = -\frac{c_{AP}}{RT}\frac{d\gamma_i}{dc_{AP}} \qquad (30)$$

This equation is the most usual approach to express the Gibbs adsorption equation. Such equation is valid for reversible transformations (dispersions in equilibrium) to constant *p* and *T*, therefore:

$$\Gamma_{AP} = -\frac{c_{AP}}{RT}\left(\frac{\partial \gamma_i}{\partial c_{AP}}\right)_{p,T,\text{reversible process}} \qquad (31)$$

### 2. Surface pressure

The surface pressure explains the interfacial tension decrease with the addition of amphiphile. The work needed to increase a surface area is applied against an unbalanced force that acts over the molecules present in the liquid surface. This force arises from resultant van der Waals interactions which are directed from surface to inside of the liquid. Interfacial tension expresses directly such parameter as described in (14). Thus, $\gamma_i$ contributes to compression of the W or O phase in droplets-based emulsions. With the adsorption of amphiphilic monomolecular layers around these droplets, a surfasse pressure ($\pi$) is generated due to lateral interactions between the polar ($\pi_W$) and nonpolar ($\pi_O$) groups of the amphiphile molecules as shown in **Figure S1**. Thus, we have:[3]

$$\pi = \pi_W + \pi_O \tag{32}$$

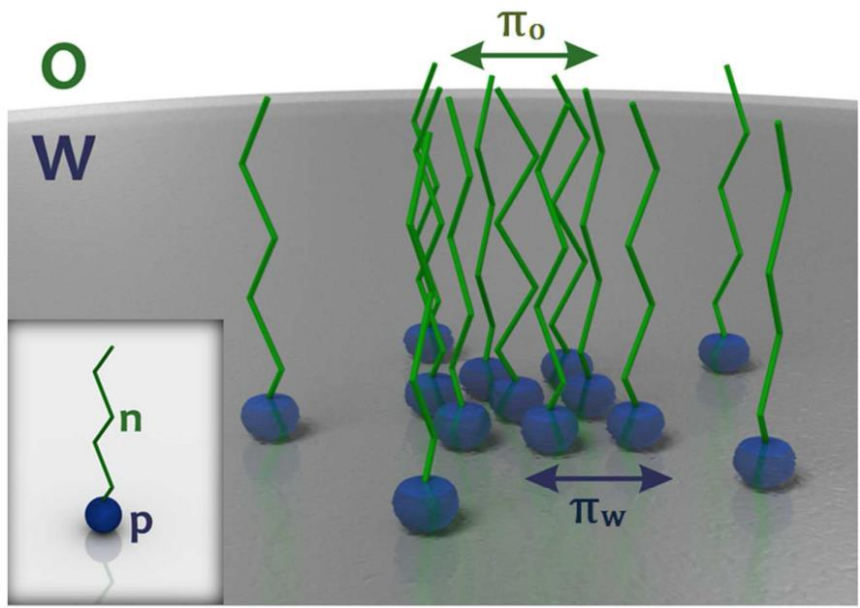

**Figure S1.** Surface activity at an interface between the water (W) and oil (O) phases. **Inset 1**, amphiphile molecule containing polar (p) and nonpolar (n) groups in the W and O phases, respectively.

In contrast with the interfacial tension, $\pi$ favors the expansion of the droplets reducing, hence, the $\gamma_i$ values as follows:

$$\gamma_i = \gamma_0 - \pi \tag{33}$$

where $\gamma_0$ is the initial interfacial tension before the amphiphile adsorption. Concluding, the formation of $\pi$ by the amphiphilic monomolecular layers around the droplets explains the reduction of $\gamma_i$ with the amphiphile content in emulsions and ME.

## 3. Microemulsification

Mixtures composed of W and O phases present interfacial tensions on the order of 30 to 50 mN m$^{-1}$. When these systems are stirred, the penetration of the phases is verified generating droplets of a phase (inner) disperses in the other (outer). Nevertheless, this process leads to an increase of the interface area raising, thus, the surface free energy as shown in (17). Therefore, such medium are thermodynamically unstable so that the droplets coalesce and the liquids separate again when the stirring is ended.[1]

In practice, the thermodynamic stability of W-O systems is ensured by reducing its interfacial tension to values of approximately 10$^{-3}$ to 10$^{-5}$ mN m$^{-1}$ through the addition of amphiphiles as discussed above. Based on (17), the decrease of $\gamma_i$ has to be more prominent or at least on the same order of magnitude concerning the raise of $\sigma$ so that $dG \leq 0$. In this situation, the emulsion or Winsor system will be thermodynamically stable generating dispersions known like microemulsions (ME). Additionally, the reduction in $\gamma_i$ has to compensate other phenomena such as (i) thermal agitation and (ii) electrostatic repulsion between the hydrophilic groups of ionic amphiphiles.[1]

## ■ Dimensional characterization of the droplet-based ME

The characterization of microemulsions taking into account diverse parameters (as surface morphology, nanostructure size and shape, structure transition, free amphiphile concentration, micellar molar composition, number of aggregation, amphiphile interfacial area, diffusion properties, relaxation behavior, viscosity, and solubility capacity) is performed through different techniques. These include electron microscopy based on cryogenic and freeze-fracture preparation, scattering techniques, nuclear magnetic resonance, spectroscopy methods, rheology, and conductivity.[4]

In this paper, the mean hydrodynamic diameter ($d_h$) of the discontinuous water-chlorobenzene ME was calculated by light scattering measurements which are based on refractive index gradients. We used the equation:

$$D = \frac{k_B T}{3\pi\eta d_h} \tag{34}$$

being $D$ the z-average diffusion coefficient, $k_B$ the Boltzmann's constant, $T$ the absolute temperature, and $\eta$ the solvent viscosity. This viscosity was determined by measuring the correlation function of the scattered intensity of tracers dispersed in the solutions (data not shown here). Silicon dioxide particles presenting 0.5 μm diameter and low size dispersity were used as tracers. The diffusion coefficient, in turn, was calculated from its relationship with the mean relaxation rate ($\bar{\Gamma}$):

$$\bar{\Gamma} = Dq^2 \tag{35}$$

with:

$$q = \frac{4\pi n}{\lambda} \sin\frac{\theta}{2} \tag{36}$$

where $q$ is the modulus of the scattering vector, $n$ is the index of refraction of the solvent, $\lambda$ is the laser wavelength, and $\theta$ is the scattering angle. The values of $\lambda$ and $\theta$ were equals to 632.8 nm and 20º, respectively. The index of refraction was determined using an Abbe refractometer (Optronics, model WYA-2S, Henan, China). Lastly, in order to calculate the mean relaxation rate, normalized intensity correlation functions, $g^{(2)}(t)$, were obtained for the dispersions. In this case, discontinuous ME (regions **A** and **C** in **Figure 1**) were investigated. Briefly, solutions composed of 52.50% v/v ethanol to water (region **A**) and 29.50% v/v ethanol to chlorobenzene (region **C**) were initially prepared. Afterwards, small amounts of the inner phase (oil in **A** and water in **C**) were added to obtain ME along the gray (**A**) and blue (**C**) dashed lines shown in **Figure 1**. Their compositions are placed in the white parts of the phase diagram, which are relative to ME (regions highlighted by red circles in **Figure 2** inset). The volume fractions of chlorobenzene added in the dispersion for region **A** ($\Phi_{O,D}$) were: 0.00%, 4.30%, 4.60%, 4.80%, and 5.10% v/v. For the region **C**, the volume fractions of transferred water ($\Phi_{W,D}$) were: 0.00%, 3.40%, 6.60%, 9.50%, 11.20%, and 12.30% v/v. The $g^{(2)}(t)$ functions were satisfactorily fitted by the cumulative expansion:

$$g^{(2)}(t) = 1 + \beta e^{-2\bar{\Gamma}t + \frac{\gamma t^2}{2}} \tag{37}$$

where $\gamma$ is a polydispersity factor, $\beta$ is a geometrical factor that depends on number of coherence areas, and $t$ is the time. Correlation functions achieved from (37) are illustrated in **Figure S2** for different volume fractions of the inner phase.

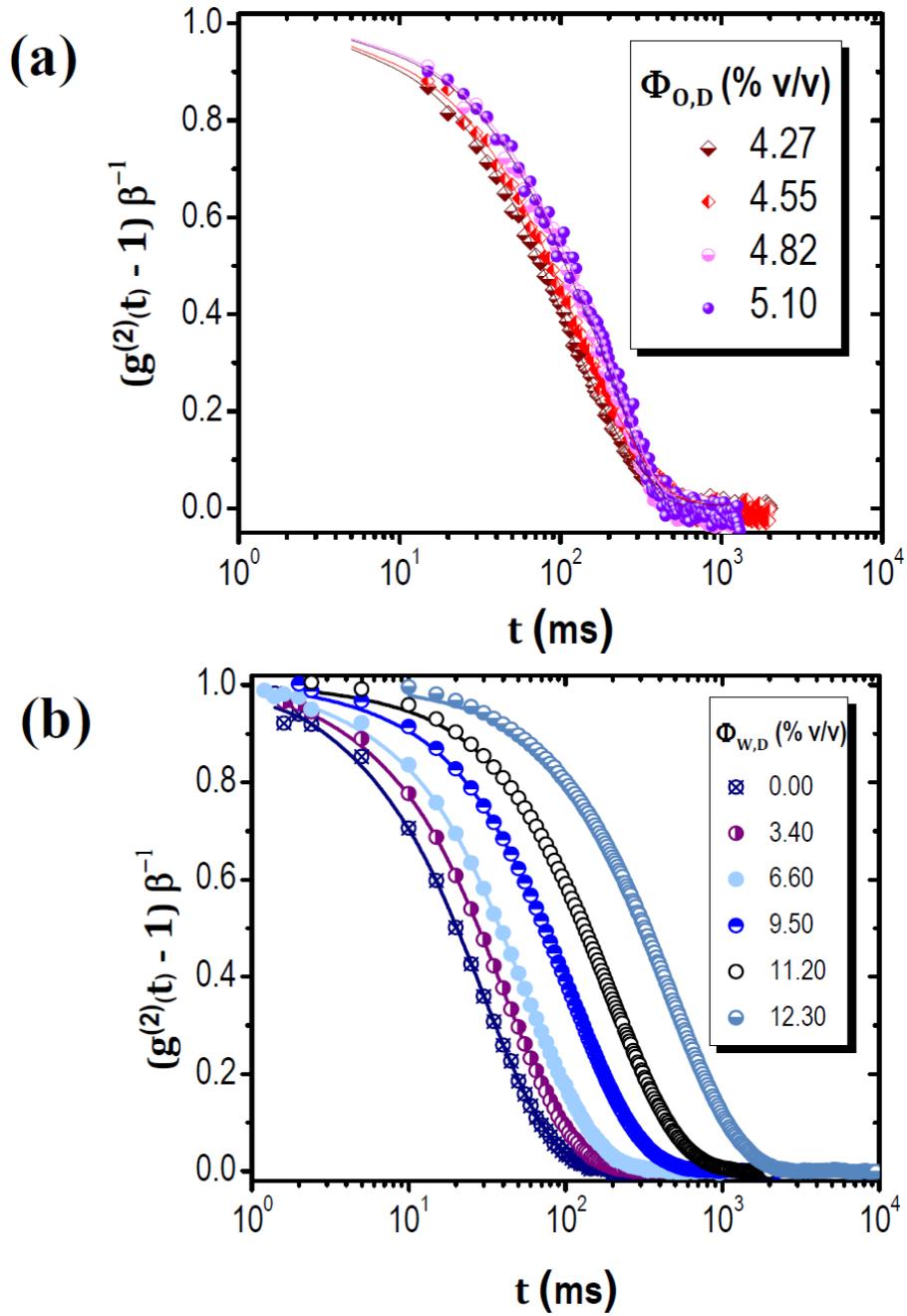

**Figure S2.** Normalized correlation functions for different concentrations of inner phase for ME prepared in the **A** (a) and **C** (b) regions of **Fig. S-2**. $\Phi_{O,D}$ and $\Phi_{W,D}$, volume fractions of chlorobenzene and water added in the total dispersion, respectively.

The mean hydrodynamic diameter values were empirically fitted to the function:

$$d_h = d_{h0} + \frac{\delta\, \Phi_{i,AP}}{\Phi_C - \Phi_{i,AP}} \tag{38}$$

where $\delta$, $d_{h0}$, and $\Phi_C$ are the fitting parameters, namely: $d_{h0}$ is the structure diameter without inner phase (0.7 ± 0.1 nm for both **A** and **C** regions), $\Phi_C$ is the critical volume fraction of this inner phase in the mixture to produce ME falling on the binodal curve, and the subscript i refers to the added oil or water. The parameter $\delta$ does not have a direct interpretation, but it can be seen as the scaling factor that connects the volume fraction with the ME's diameter. The values of $\Phi_C$ for **A** and **C** were 5.4 ± 0.4% and 13.1 ± 0.1% v/v, respectively. The obtained $d_h$ values are shown in **Figure 2**.

## ■ Reasons for differences in sensitivity

For the different sensitivities in the tested regions of the water-chlorobenzene ME, our hypothesis relates to the effect of the AP-added water over the microemulsification process by changing the surface area and pressure.

As discussed in the main text, the positive angular coefficients observed in the analytical curves are because the decrease in content of ethanol AP as the volume fraction of water to ethanol ($\Phi_W$, analyte concentration) increases. The displacement of the phase behavior for each investigated region with the addition of water in ethanol is shown by arrows in **Figure 1**. In **A**, gradual additions of water increase the efficiency of the ME so that a lower volume of amphiphile is required to stabilize the media. Thereby, the positive deviations in the MEC response ($\Phi_{ME}$, minimum volume fraction of amphiphile needed to form ME) with $\Phi_W$ are attenuated decreasing the levels of sensitivity in MEC. Such result likely arose from the increase in hydration of the polar groups of the AP molecules. It raises the $\pi_W$, thus increasing the efficiency of reduction of $\gamma_i$ by the AP.[24] This contributes for thermodynamic stabilization of the dispersions by decreasing $G$, what requires a lower value of $\Phi_{ME}$. For the data in region **C**, we verified an opposite phenomenon. Herein, the successive additions of water reduce the efficiency. Consequently, the modifications in $\Phi_{ME}$ with $\Phi_W$ are increased which improve the sensitivity. The theoretical reasons for such behavior are likely related to a larger $\sigma$ due to the increase in inner phase fraction. This requires a further reduction in $G$ by decreasing $\gamma_i$ through the addition of ethanol.[1] Lastly, the $\Phi_{ME}$ in region **B** remains almost invariable with the additions of water because a reduction of the aforesaid phenomena. Thus, the assays in **B** had sensitivity intermediate in relation to **A** and **C**.

**Figure S3** displays the magnitude of variation of the MEC signal, $\Phi_{ME}$, with the volume fractions of water to chlorobenzene ($\Phi_{W,O}$) in **A**, **B**, and **C**. Despite the poor quality of the linear fitting, the angular coefficients can be used to evaluate such variation. Thus, the effect of the water over $\Phi_{ME}$ in **B** (0.09) is very low, whereas this is higher in **A** (-1.49, decrease in $G$ with water content) and **C** (1.72, increase in $G$ with water content).

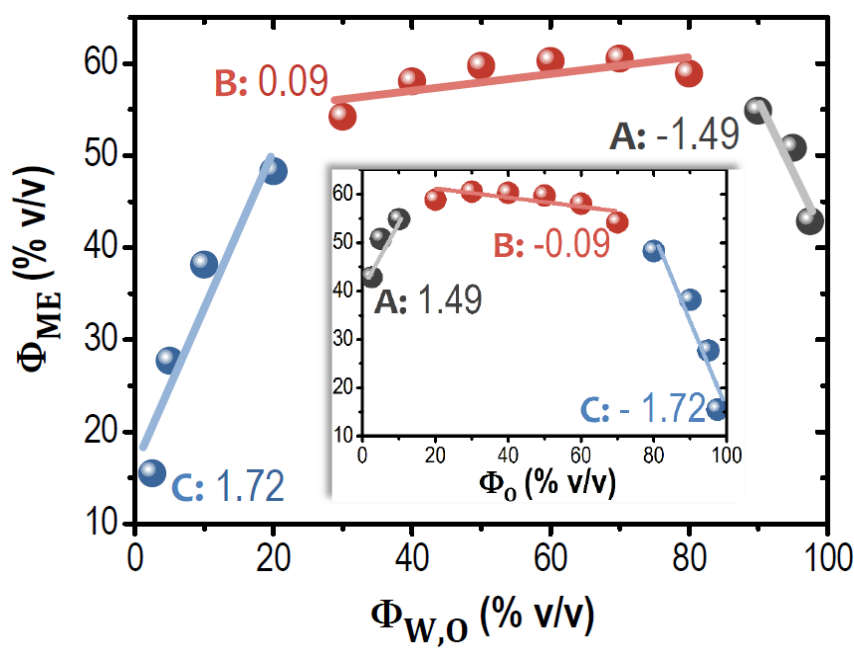

**Figure S3.** Variation of the MEC analytical response, $\Phi_{ME}$, with $\Phi_{W,O}$. $R_2$ values: 0.9305 (**A**), 0.9207 (**B**), and 0.9468 (**C**). **Inset**, variation of $\Phi_{ME}$ with the volume fractions of chlorobenzene to water ($\Phi_O$). Angular coefficients for each case are shown.

As shown in **Figure S4**, the limits of linearity (**LOL**) in **B** and **C** were in agreement with our hypothesis that takes into account the AP-added water to explain the different sensitivity and detectability obtained for the inspected regions. This figure expresses the compositions related to all of the analytical curve points in phase diagrams. Based on such, we observed that the **LOL** for **C** is found in region **B**, where the water does no contribute more to modify $\Phi_{ME}$. For **B**, in turn, the **LOL** value is in **A**. In this region, the water has an appreciable effect over the microemulsification by reducing $G$ and, therefore, $\Phi_{ME}$. Conversely, the causes for the **LOL** as well as the two linear ranges in region **A** are unknown yet.

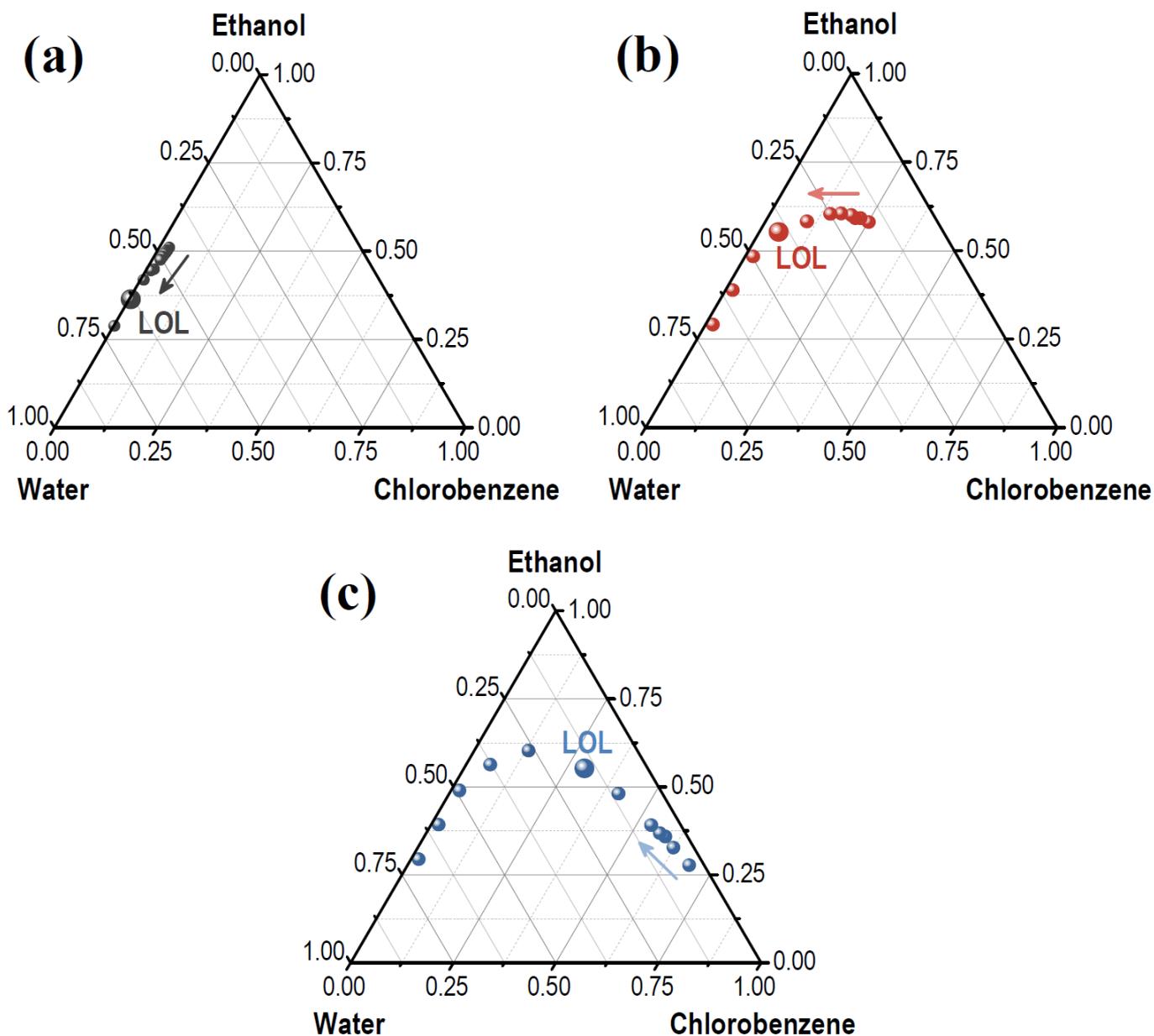

**Figure S4.** Phase diagrams containing the points of the analytical curves for the **A** (a), **B** (b), and **C** regions (c). The values of **LOL** are depicted in each tested region.

Concerning the difference of sensitivity in region **B** of the diagram for application to monoethylene glycol, the effect of the water on microemulsification is presumably the reason again for such phenomenon as stated in the main text. The negative angular coefficients that were observed in the analytical curve are due to the decrease in interface tension as $\Phi_M$ (fraction of monoethylene glycol to water, analyte content) increases. Concomitantly, it represents a reduction in the volume fraction of water to O phase ($\Phi_{W,O}$), which has also effect over the $\gamma_i$. The two linear ranges of the analytical curve are related to regions where the change magnitudes of $\Phi_{ME}$ with $\Phi_{W,O}$ are distinct as illustrated in **Figure S5**.

Below about 40.00% v/v $\Phi_{W,O}$, the reduction slope of $\Phi_{ME}$ with $\Phi_{W,O}$ is increased from 18.86 to 145.48. It indicates a higher intensity of decrease in $\gamma_i$. Thereby, it is expected that the negative deviation of $\Phi_{ME}$ in the analytical curve raises for values of $\Phi_M$ greater than approximately 60.00% v/v. This was observed, improving the sensitivity in the second part of the analytical curve. Herein, the **S** increased from 0.20 to 0.44 for $\Phi_M$ larger than 60.00% v/v.

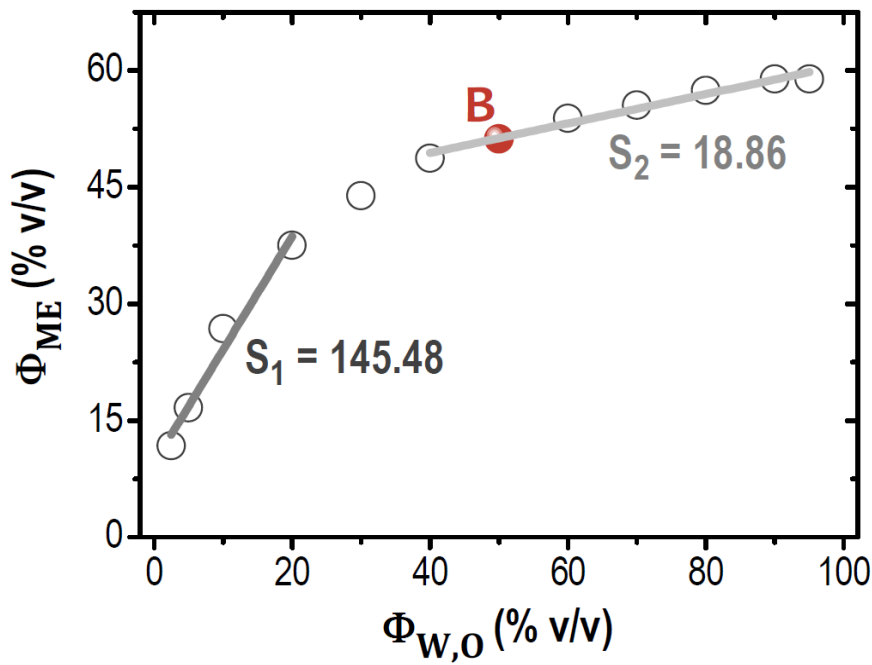

**Figure S5.** Variation of MEC analytical response with $\Phi_{W,O}$. The values of $\Phi_{ME}$ are related to the binodal curve in **Figure 5** ($\Phi_M$ = 0.00) of the main text. $R_2$ values: 0.9861 and 0.9881 for first and second linear range, respectively.

## ■ Temperature-function robustness

Regarding the investigation about the robustness of the MEC as function of changes in temperature, the resulting $\Delta\Phi$ values (absolute error determined for $\Phi_W$) for the changes of 23 to 26 °C are shown in **Figure S6**. In general, $\Delta\Phi$ ranged from approximately 0.30 to 5.00% v/v. The data attained in region **C** presented the best robustness level.

## ■ Characterization of the natural gas samples

The liquefied natural gas samples were provided by the research center (*Centro de Pesquisas e Desenvolvimento Leopoldo Américo Miguez de Mello*) of Petrobras. These samples were characterized in relation to the presence of anions, carboxylic acids, metals, and silicon through different instrumental analytical methods. The metals and silicon were determined using inductively coupled plasma atomic emission spectroscopy (Perkin Elmer-Sciex, model Optima 3300 DV, Waltham, MA). The anions carboxylic acids, in turn, were analyzed through ion chromatography (Metrohm 861 Advanced Compact IC, model MSM II, Herisau, Switzerland).

The obtained results are shown in **Table S1**.

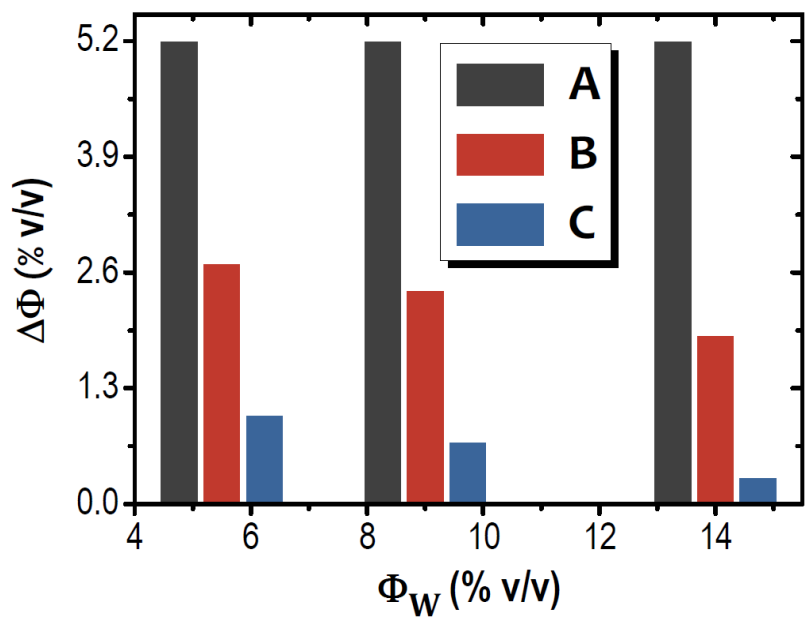

**Figure S6.** Values of ΔΦ as function of $\Phi_w$ in three regions of the analytical curves (5.50%, 9.00%, and 14.00% v/v $\Phi_w$) for the change of 23 to 26 °C in the **A**, **B**, and **C** regions. The ΔΦ parameter is given in module; some of its obtained values were negative.

**Table S1.** Concentrations of diverse compounds present in the liquefied natural gas samples (**R1-R4**). The sample **R4** was not analyzed concerning the presence of anions and carboxylic acids. N.D. means non-detected.

| Samples | Anions (mg L$^{-1}$) | | | Carboxylic Acids (mg L$^{-1}$) | | | | | Metals (mg L$^{-1}$) | | | | | | | Si (mg L$^{-1}$) |
|---|---|---|---|---|---|---|---|---|---|---|---|---|---|---|---|---|
| | Cl$^-$ | Br$^-$ | SO$_4^{2-}$ | Glycolate | Formate | Acetate | Propionate | Butyrate | Na$^+$ | K$^+$ | Mg$^{2+}$ | Ca$^{2+}$ | Ba$^{2+}$ | Sr$^{2+}$ | Fe$^{3+}$ | |
| **R1** | 10 | 15 | 230 | N.D. | 20 | 75 | 10 | < 5 | 2 | < 1 | < 1 | 1 | < 1 | < 1 | 1 | < 1 |
| **R2** | 5,100 | 150 | 170 | 75 | 15 | 45 | < 5 | < 5 | 1,600 | 140 | 87 | 270 | 2 | 22 | 52 | 2 |
| **R3** | 850 | 110 | 310 | 70 | 7 | 44 | N.D. | N.D. | 398 | 38 | 14 | 77 | 1 | 9 | 8 | 2 |
| **R4** | - | - | - | - | - | - | - | - | 27 | 8 | < 1 | 9 | < 1 | 1 | < 1 | < 1 |

# ■ References